\documentclass[sigconf]{acmart}
\makeatletter
\def\@ACM@checkaffil{
    \if@ACM@instpresent\else
    \ClassWarningNoLine{\@classname}{No institution present for an affiliation}%
    \fi
    \if@ACM@citypresent\else
    \ClassWarningNoLine{\@classname}{No city present for an affiliation}%
    \fi
    \if@ACM@countrypresent\else
        \ClassWarningNoLine{\@classname}{No country present for an affiliation}%
    \fi
}
\makeatother
\usepackage[utf8]{inputenc}

\usepackage[english]{babel}
\usepackage{blindtext}
\usepackage{color}
\usepackage{xspace}
\usepackage{caption}
\usepackage{multirow}
\usepackage{amsmath}  
\usepackage{bm}
\usepackage{graphicx}
\usepackage{gensymb}
\usepackage{booktabs}
\usepackage{algorithm}
\usepackage{algpseudocode}
\usepackage{subcaption}
\usepackage[compact]{titlesec}
\usepackage{tikz}
\usepackage{xcolor}

\copyrightyear{2024}
\acmYear{2024}
\setcopyright{rightsretained}
\acmConference[IMC '24]{Proceedings of the 2024 ACM Internet Measurement Conference}{November 4--6, 2024}{Madrid, Spain}
\acmBooktitle{Proceedings of the 2024 ACM Internet Measurement Conference (IMC '24), November 4--6, 2024, Madrid, Spain}\acmDOI{10.1145/3646547.3689006}
\acmISBN{979-8-4007-0592-2/24/11}

\begin{document}

\newcommand{\BULLET}{\vspace{+.00in} \noindent $\bullet$ \hspace{+.00in}}

\newcommand{\etc}{\emph{etc.}\xspace}
\newcommand{\ie}{\emph{i.e.,}\xspace}
\newcommand{\eg}{\emph{e.g.,}\xspace}
\newcommand{\etal}{\emph{et al.}\xspace}
\newcommand{\wrt}{\emph{w.r.t.}\xspace}
\newcommand{\aka}{\emph{a.k.a.}\xspace}

\newcommand{\DEG}{\degree\xspace}
\newcommand{\accessdate}{09/11/2024}

\newcommand*\circled[1]{\tikz[baseline=(char.base)]{
            \node[shape=circle,draw,inner sep=2pt] (char) {#1};}}
            
\newcommand{\name}{$\sf\small{Theia}$\xspace}	
\newcommand{\boldname}{\textbf{Theia}\xspace}
\newcommand{\bigname}{Theia\xspace}

\newcommand{\hd}[1]{\small{\textbf{\texttt{#1}}}\normalsize}
\newcommand{\red}[1]{\textcolor{red}{#1}}
\newcommand{\blue}[1]{\textcolor{blue}{#1}}

\newcounter{RZNumberOfComments}
\stepcounter{RZNumberOfComments}
\newcommand{\rz}[1]{\textcolor{teal}{\small \bf [RZ\#\arabic{RZNumberOfComments}\stepcounter{RZNumberOfComments}: #1]}}

\newcounter{BHNumberOfComments}
\stepcounter{BHNumberOfComments}
\newcommand{\bo}[1]{\textcolor{red}{\small \bf [BH\#\arabic{BHNumberOfComments}\stepcounter{BHNumberOfComments}: #1]}}

\newcounter{MVNumberOfComments}
\stepcounter{MVNumberOfComments}
\newcommand{\mvnote}[1]{\textcolor{blue}{\small \bf [MV\#\arabic{MVNumberOfComments}\stepcounter{MVNumberOfComments}: #1]}}

\newcounter{CSQNumberOfComments}
\stepcounter{CSQNumberOfComments}
\newcommand{\csq}[1]{\textcolor{green}{\small \bf [CSQ\#\arabic{CSQNumberOfComments}\stepcounter{CSQNumberOfComments}: #1]}}

\newcommand{\nan}[1]{{\color{blue}#1}}

\title{ 
A First Look at Immersive Telepresence on Apple Vision Pro
}

\settopmatter{authorsperrow=3}

\author{Ruizhi Cheng$^{*}$}
\affiliation{
  \institution{George Mason University}
  }
\email{rcheng4@gmu.edu}

\author{Nan Wu$^{*}$}
\affiliation{
  \institution{George Mason University}
  }
\email{nwu5@gmu.edu}

\author{Matteo Varvello}
\affiliation{
  \institution{Nokia~Bell Labs}
  }
\email{matteo.varvello@nokia.com}

\author{Eugene Chai}
\affiliation{
  \institution{Nokia~Bell Labs}
  }

\email{eugene.chai@nokia-bell-labs.com}

\author{Songqing Chen}
\affiliation{
  \institution{George Mason University}
  }
\email{sqchen@gmu.edu}

\author{Bo Han}
\affiliation{
  \institution{George Mason University}
  }
\email{bohan@gmu.edu}

\begin{abstract}
Due to the widespread adoption of ``work-from-home'' policies, videoconferencing applications {(\eg Zoom)} have become indispensable for remote communication.
However, they often lack immersiveness, leading to the so-called ``Zoom fatigue'' and degrading communication efficiency. 
The recent debut of Apple Vision Pro, a mobile headset that supports ``spatial persona'', aims to offer an immersive telepresence experience.
In this paper, we conduct a first-of-its-kind in-depth and empirical study to analyze the performance of immersive telepresence with Apple FaceTime, Cisco Webex, Microsoft Teams, and Zoom on Vision Pro.
We find that only FaceTime provides a truly immersive experience with spatial personas, whereas others still operate 2D personas.
Our measurement results reveal that (1) FaceTime delivers semantic data to optimize bandwidth consumption, which is even lower than that of 2D persona for other applications, and (2) it employs visibility-aware optimizations to reduce rendering overhead.
However, the scalability of FaceTime remains limited, with a simple {server-allocation} strategy that potentially leads to high network delay for users.

\end{abstract}



\maketitle

\section{Introduction}
\label{sec:intro}
Remote communication is indispensable in contemporary life, even in the post-pandemic era, as evidenced by $\sim$90\% of meetings involving remote participants in 2024~\cite{meeting_statistics}.
Existing remote communication systems predominantly rely on traditional 2D video-based conferencing. 
These platforms often lack the ability to convey social signals such as eye contact and body language, leading to inefficient communication~\cite{neate2022just} and so-called ``Zoom fatigue''~\cite{zoomfatigue}.

\def\thefootnote{*}\footnotetext{These authors contributed equally to this work.}
\def\thefootnote{\arabic{footnote}}

\begin{figure}[t]
    \centering   
    \includegraphics[width=0.36\columnwidth]{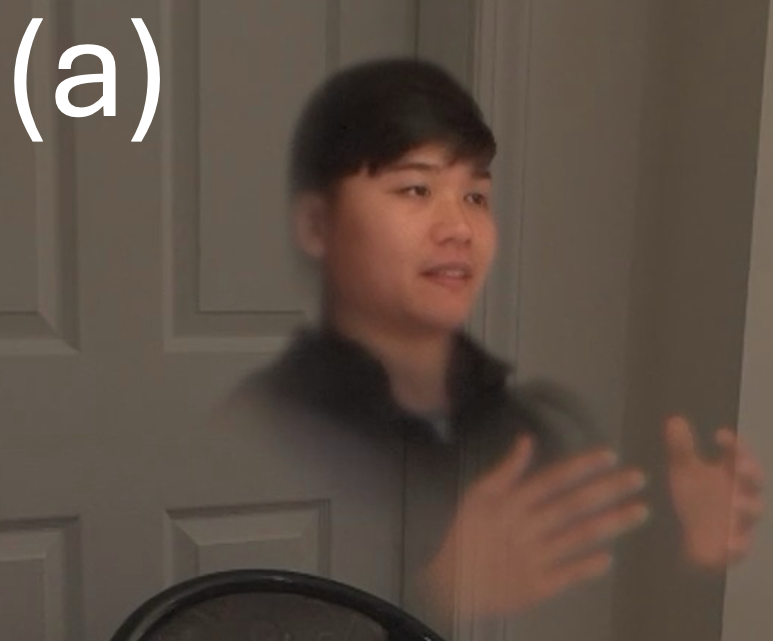} 
    \includegraphics[width=0.36\columnwidth]{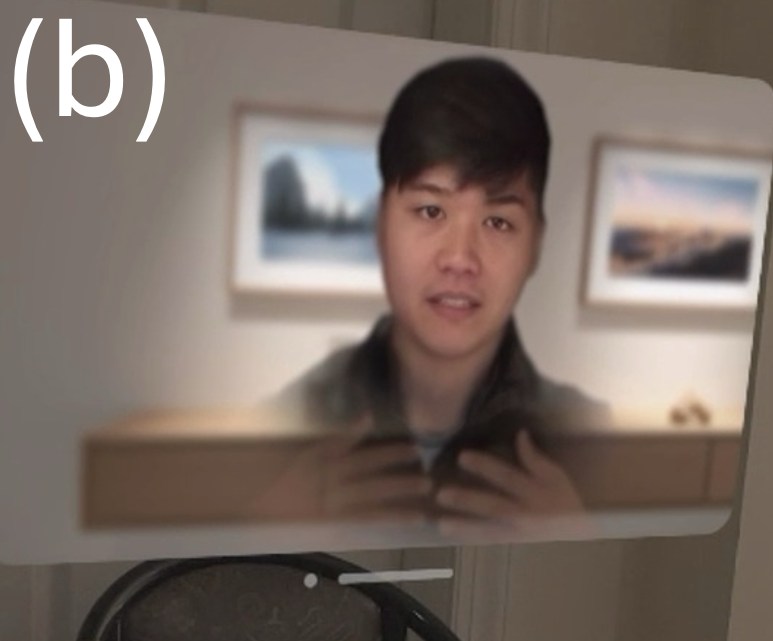} 
    
    \vspace{-0.05in}
    \caption{(a) Spatial persona on FaceTime \textit{vs.} (b) 2D persona on Webex.} 
    \label{fig:persona}
    \vspace{-0.2in}
\end{figure}

Immersive telepresence is a game changer in remote communication by offering engaging and interactive experiences and is widely recognized as the top use case in the forthcoming 6G~\cite{tariq2020speculative,de2021survey}.
Despite its promise, the commercial availability of immersive telepresence systems has been limited.
Although several tech giants have launched a few projects on immersive telepresence~\cite{lawrence2021project,ma2021pixel,orts2016holoportation}, with arguably the earliest one dating back to 2016~\cite{orts2016holoportation}, they largely remain internal endeavors with almost no public access. 
Meanwhile, academic research in this area typically focuses on in-lab prototypes~\cite{guan2023metastream,jin2024meshreduce,ihara2023holobots}.
The recent debut of Apple Vision Pro~\cite{visionpro}, a mixed reality (MR) headset that supports ``spatial persona'', as shown in Figure~\ref{fig:persona} and introduced in \S\ref{sec:bakcground}, marks a significant milestone in immersive telepresence. 
Vision Pro allows users to pre-capture their personas, which are 3D human models capable of tracking their hand and head movements in real time.

In this paper, we conduct, to the best of our knowledge, the first measurement study to dissect the functioning and performance of immersive telepresence, focusing on four videoconferencing applications (VCAs) for Vision Pro: Apple FaceTime~\cite{facetime}, Cisco Webex~\cite{webex}, Microsoft Teams~\cite{teams}, and Zoom~\cite{zoom}.
We summarize our key findings as follows. 

\vspace{0.02in}
\BULLET All VCAs assign a server 
near the initiating user of a telepresence session, potentially leading to $>$100~ms network delays even when all users are located in the US.

\vspace{0.02in}
\BULLET Only FaceTime offers a truly immersive telepresence experience with spatial personas. 
{Moreover}, its bandwidth consumption~($<$0.7 Mbps) is {even} lower than other platforms that deliver 2D personas~(\eg $>$4 Mbps on Webex). 
The reason is that FaceTime 
benefits from emerging semantic communication~\cite{cheng2023enriching}, instead of directly streaming 3D content or 2D video. 

\vspace{0.02in}
\BULLET {The delivery of a spatial persona does not support rate adaption, mainly due to its employment of semantic communication. 
{This is because semantic communication requires all semantic data to be fully delivered for accurate reconstruction, making it challenging to adapt to varying bandwidth conditions~\cite{cheng2023enriching}.

\vspace{0.02in}
\BULLET Spatial personas on FaceTime leverage visibility-aware optimizations~\cite{han2020vivo} to decrease rendering time by up to 59\%. Yet, these optimizations are not exploited to reduce bandwidth consumption. 

\vspace{0.02in}
\BULLET The scalability of FaceTime remains limited. 
As the number of users grows, its CPU/GPU processing time increases correspondingly, and the bandwidth consumption rises almost linearly. 
{The GPU processing time reaches $\sim$9 ms per frame when there are five users, 
close to the 11.1 ms requirement for 90 frames per second~(FPS) rendering on Vision Pro~\cite{realitykit}.
This explains why FaceTime currently supports a maximum of only five spatial personas~\cite{spatial_persona}.
}

Our findings contribute to a comprehensive understanding of the current design and development of immersive telepresence systems and their performance bottlenecks.
The source code and data used in this paper are available at \url{https://github.com/felixshing/IMC2024VisionPro}.
This work has been approved by the institutional review board (IRB) and does not raise any ethical issues.

\section{Background}
\label{sec:bakcground}

\noindent{\bf Spatial Persona \textit{vs.} 2D Persona.}
In immersive telepresence, a persona is a dynamic digital representation of a participant that facilitates interactions with others. 
Apple Vision Pro's personas capture users' face, hand, and eye movements to make remote communication engaging.
It offers two representations: spatial personas and 2D personas.
Figure~\ref{fig:persona}(a) shows the spatial persona on FaceTime.
It can be viewed from different angles in real time, providing an immersive and interactive experience.
As of the time of our measurement study (April 2024), we found that the spatial persona is available on only FaceTime.
In contrast, the personas on other applications are still 2D, as shown in 
Figure~\ref{fig:persona}(b) for Webex.
It is generated for a static viewport, functioning as if recorded by a virtual camera in these applications that mimics the selfie camera.
This means when a user moves, 
the display of remote participants' 2D personas does not 
change accordingly.

\begin{figure}[t]
    \centering    \includegraphics[width=0.45\columnwidth]{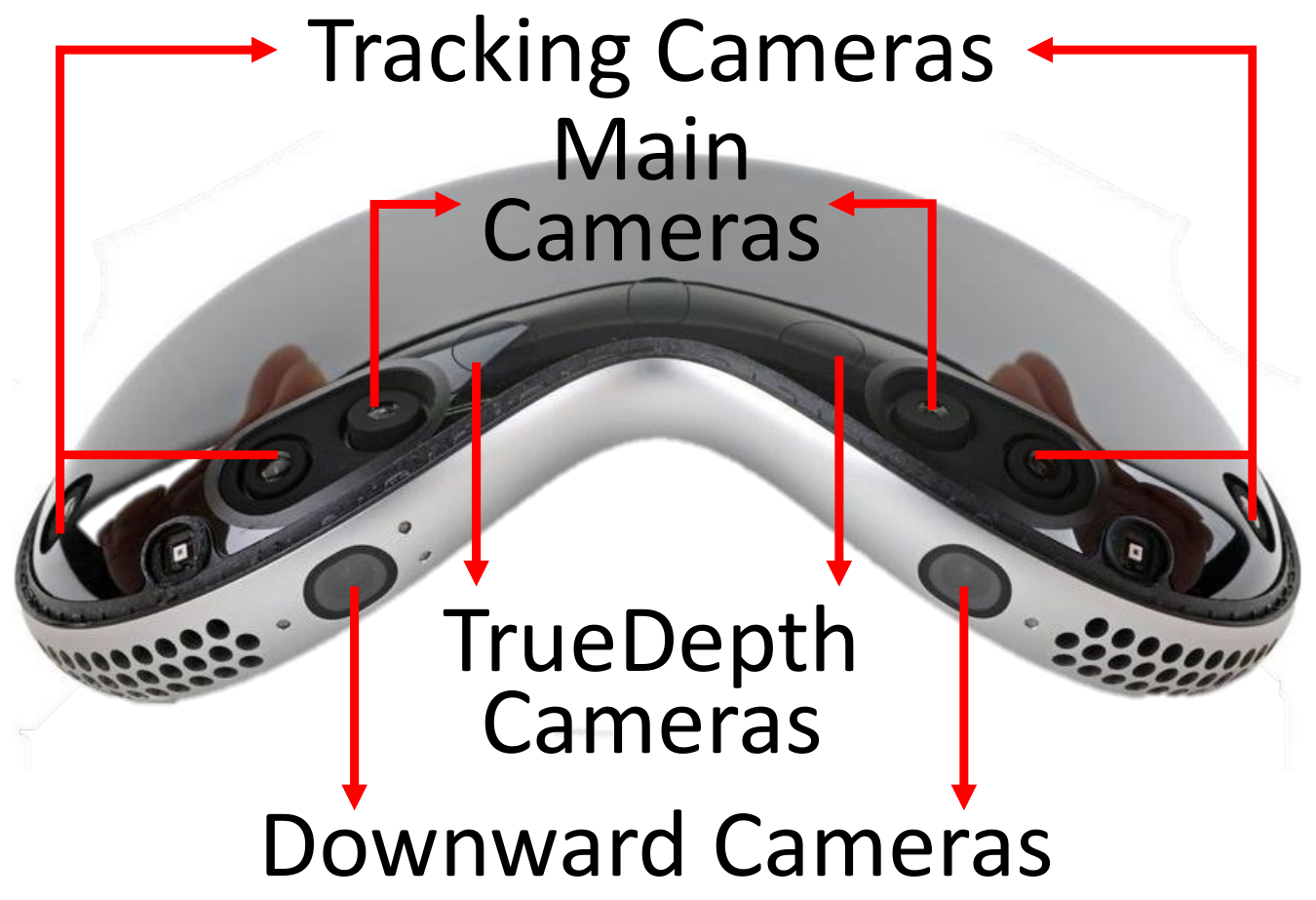} 
    \includegraphics[width=0.36\columnwidth]{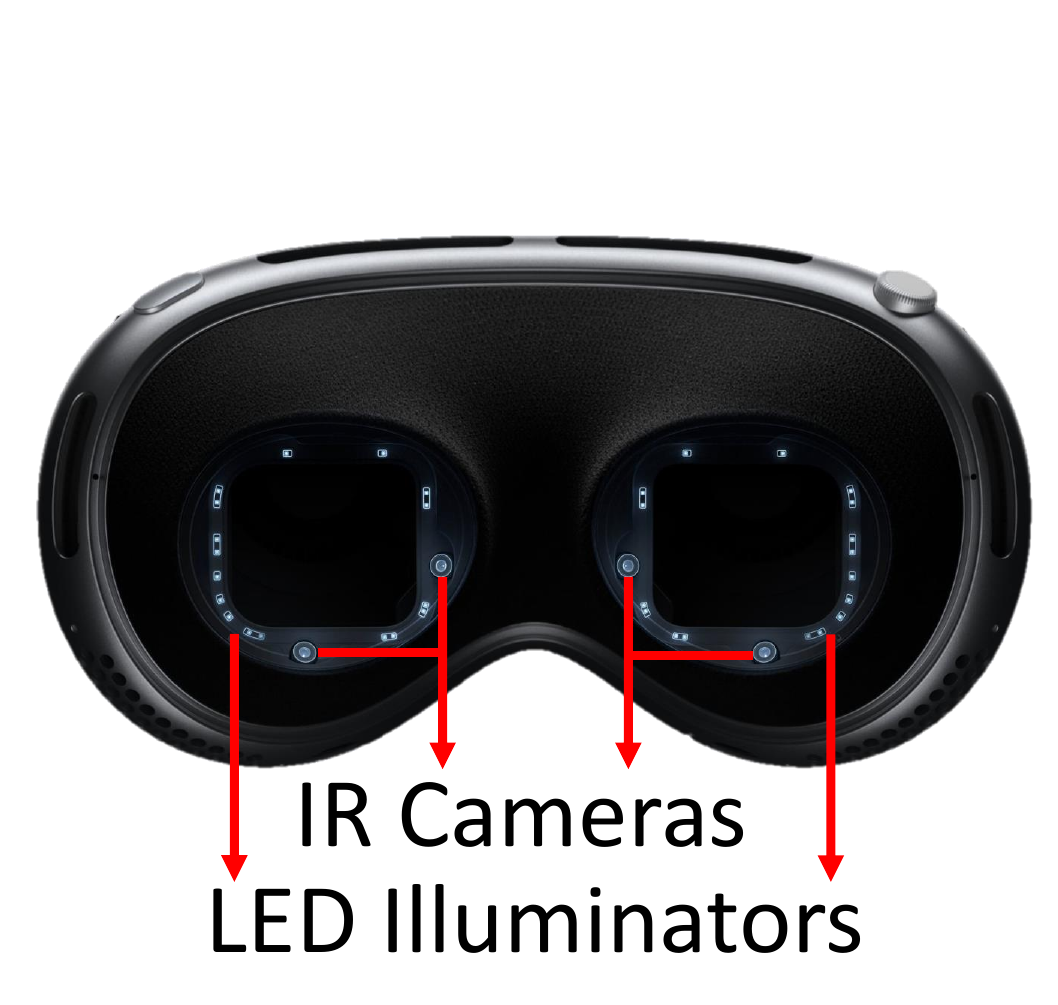} 
    \vspace{-0.05in}
    \caption{Cameras on Apple Vision Pro.} 
    \label{fig:visionpro}
    \vspace{-0.2in}
\end{figure}

\vspace{0.02in}
\noindent{\bf Mobile MR Headsets}
blend digital content with the real world, offering interactive experiences that bridge virtual and physical spaces.
Optical see-through devices, such as Microsoft HoloLens~2~\cite{hololens2} and MagicLeap 2~\cite{magicleap2}, allow users to directly view their environment with digital overlays projected via transparent lenses.
On the other hand, video see-through headsets, such as Meta Quest 3~\cite{quest3} and Apple Vision Pro~\cite{visionpro}, capture the surrounding environment through their cameras and then display the imagery combining digital and real-world content on their screens. 
Figure~\ref{fig:visionpro} shows the cameras on Apple Vision Pro. 
The main cameras on the front provide a see-through view of the real world, and the tracking cameras {sense the user's position and neighboring objects.}
{The \textit{TrueDepth} cameras can be used to pre-capture the spatial persona offline}, and the downward cameras monitor the user's face.
Additionally, the {internal IR cameras} 
track the user's eyes to offer {better experiences}, such as enabling eye contact 
in immersive telepresence.

\section{Experimental Setup}
\label{sec:setup}
\begin{figure}[t]
    \centering    \includegraphics[width=0.95\columnwidth]{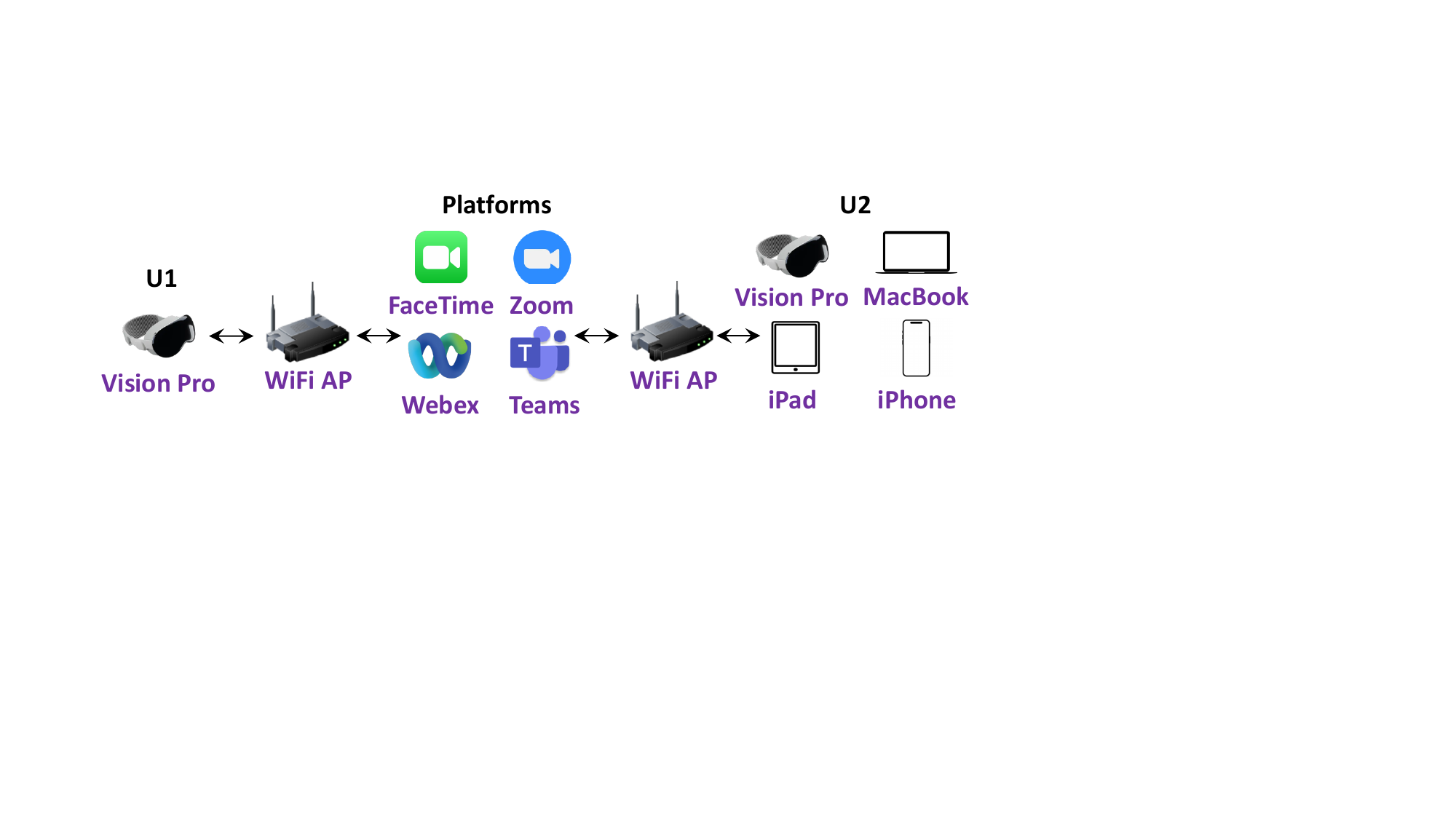} 
    \vspace{-0.1in}
    \caption{Measurement setup with two users, U1 and U2.}     
    \label{fig:setup}
    \vspace{-0.15in}
\end{figure}

In this section, we describe the VCAs under investigation, the testbed setup, and the performance metrics of our measurement experiments conducted in April 2024.

\subsection{Videoconferencing Applications}
We investigate four popular VCAs: Apple FaceTime~\cite{facetime}, Cisco Webex~\cite{webex}, Microsoft Teams~\cite{teams}, and Zoom~\cite{zoom}.
We choose Apple FaceTime because it 
supports spatial personas~\cite{spatial_persona}, enabling an immersive experience for Vision Pro users.
%
The other three applications have been extensively studied by the research community~\cite{chang2021can, MacMillan2021VCAs, michel2022enabling, varvello2022performance, sharma2023estimating, he2023measurement} and are available on Vision Pro.

\subsection{Testbed \& Data Collection}
\label{sec:testbed}
Figure~\ref{fig:setup} shows our experimental setup.
Unless otherwise mentioned, our experiments involve two users, U1 and U2.
U1 is always equipped with Vision Pro, whereas U2 uses Vision Pro, MacBook, iPad, or iPhone. 
{Most experiments are conducted with both users wearing Apple Vision Pro. U2 uses other devices when we test traditional 2D video calls on FaceTime for the protocol (\S\ref{sec:server_infra}) and throughput (\S\ref{sec:tput}) analysis.}
%
All devices are updated to the latest version of their operating system.
U1 and U2 are connected to two different WiFi access points (APs), each with an average throughput of more than 300~Mbps.
We use Wireshark~\cite{Wireshark} on each AP to capture and analyze network traffic.
{
To assess the performance and resource utilization of Vision Pro, we use Xcode~\cite{xcode} to pair it with a dedicated MacBook where we run Apple's RealityKit tool~\cite{realitykit}.}

Similar to a prior 
study~\cite{michel2022enabling}, we collect telepresence statistics using the tools provided by Zoom~\cite{zoom_statistics}, Webex~\cite{webex_statistics}, and Teams~\cite{teams_statistics}. 
We measure network latency by running TCP \texttt{pings}~\cite{tcptraceroute} between our WiFi APs and Apple servers for FaceTime, since the servers block regular ICMP pings. 
We verify that no background process exists on the devices during our experiments.
{As our measured platforms are primarily designed for video conferencing, users in our experiments are instructed to engage in natural conversations and movements, simulating a typical meeting environment.}
{We repeat each experiment at least five times, and each session lasts at least 120 seconds.} In the following, we describe the performance metrics that we study. 

\BULLET \textit{Throughput:} We measure the throughput of these applications involving up to five participants, which is the maximum number of supported spatial personas on Vision Pro~\cite{spatial_persona}. 

\BULLET \textit{Display Latency:} 
{
We measure the difference in display latency between rendering real-world objects and the spatial personas of remote users.}
Recall that Vision Pro is a video see-through headset.
It captures and renders real-world content, and then integrates it with the rendered spatial persona~(\S\ref{sec:bakcground}).
Thus, we can record the content displayed on the headset to measure this latency.

\BULLET \textit{Frame Rate and Rendering Time for Each Frame}: 
The target FPS of Vision Pro is 90~\cite{realitykit}. 
We measure CPU/GPU processing time for each frame to identify bottlenecks if a frame misses its deadline. 

\BULLET  \textit{Visual Quality}: 
On Vision Pro, the 3D model of a spatial persona is represented as mesh~\cite{peng2005survey}.
The visual quality of a mesh is influenced by the number of triangles, which are connected to form the geometry of the 3D model.
%
For 2D personas, we resort to measuring the video resolution as done in previous work~\cite{chang2021can, MacMillan2021VCAs, michel2022enabling, varvello2022performance, sharma2023estimating}. 
For both metrics, the higher they are, the higher the rendering overhead, and {the better the} visual quality.

\section{Measurement Results}
\label{sec:results}
\subsection{Server Infrastructure}
\label{sec:server_infra}

\begin{figure}[t]
    \centering    \includegraphics[width=0.8\columnwidth]{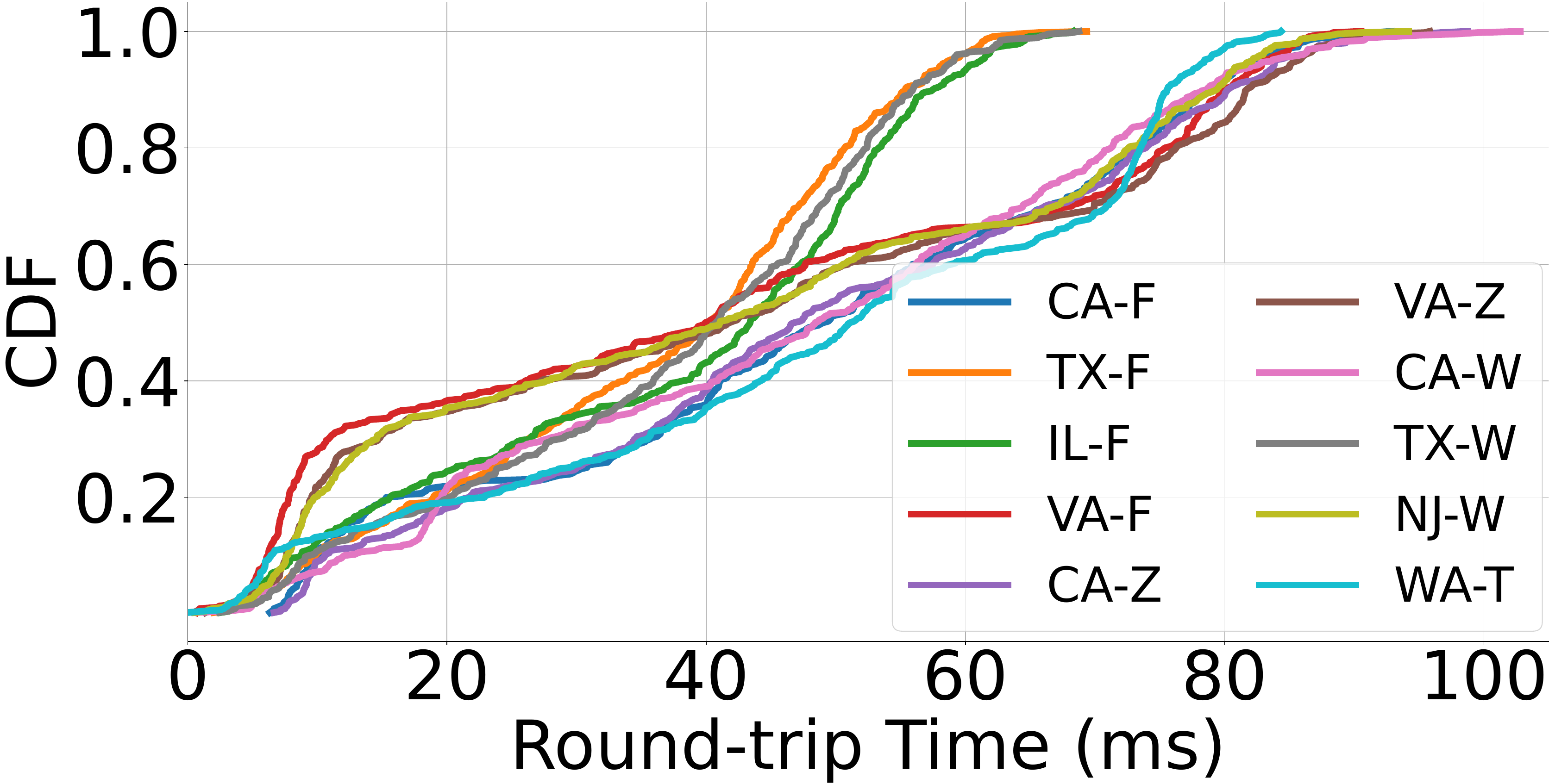} 
    \vspace{-0.1in}
    \caption{Round-trip time between FaceTime (F), Zoom (Z), Webex (W), and Teams (T) servers and test users. The server locations are indicated by their abbreviations: CA (California), TX (Texas), IL (Illinois), VA (Virginia), NJ (New Jersey), and WA (Washington State).}  
    \label{fig:cdf_rtt}
    \vspace{-0.2in}
\end{figure}

\noindent {\bf Geolocation.} %
We first investigate the server locations and network latency of the four VCAs. 
Since Vision Pro is available in only the US as of April 2024, we set up clients in nine different locations across the Western (three), Middle (three), and Eastern (three) US.
For each experiment, these {nine} clients randomly join a VCA in different orders with different types of device.
We use MaxMind~\cite{MaxMind} and ipinfo.io~\cite{ipinfo} to geolocate the servers we identify. 
{Both tools return the same geolocation results for all tested servers}.

We find that FaceTime, Zoom, Webex, and Teams operate four~{(Virginia, Illinois, California, and Texas)}, two~{(Virginia and California)}, three~{(New Jersey, California, and Texas)}, and one~{(Washington State)} server(s) in the US, respectively.
Zoom and FaceTime rely on {peer-to-peer (P2P)} communication, with data transmitted directly between users without involving a server, when there are only two users in a session, except for both users using Vision Pro on FaceTime.
We ascertain that none of the servers employ \textit{anycast}~\cite{rfc1546} by using the approach adopted by prior work~\cite{cheng2022are}.
All VCAs consistently assign a server that is closest to the initiating user of each telepresence session. 
For example, if a user in the Eastern US initiates a session, the server will always be 
in the Eastern US {(if available)}, regardless of the locations of other participants.

{Figure~\ref{fig:cdf_rtt} presents the round-trip time (RTT) between the servers of the four VCAs and our test clients. 
We observe that even though all users and servers are located within the US, the RTT between them can still exceed 100~ms, as observed with the California server of Webex.
For servers situated on the west coast (California and Washington
State) and the east coast (Virginia and New Jersey), the RTT can be  $>$80~ms when users are located on the opposite coast.
Positioning servers in the middle of the US, such as Texas and Illinois, can potentially reduce the maximum RTT, with all RTTs falling below 70 ms.
This is because the central location ensures a shorter distance to both coasts compared to the distance between the east and west coasts.
However, this strategy may decrease the percentage of clients experiencing low RTTs, given that the majority of the US population resides on the east and west coasts~\cite{census}, and the lowest RTTs occur when servers are located nearby.
For example, only 20\% of RTTs for the Texas server of FaceTime are below 20 ms, compared to 38\% for its Virginia server.

\vspace{0.02in}
{\noindent{\bf Implications~\circled{1}~:}
The above results reveal that the straightforward solution of allocating a single server for all users can result in high network latency.
This issue could become more pronounced when users are distributed across continents. 
For example, the one-way propagation delay between Europe and Asia may already exceed 100 ms~\cite{global_ping}, the threshold for maintaining a high quality of
experience (QoE) in immersive telepresence~\cite{lee2023farfetchfusion}.
A viable solution would be to deploy geo-distributed servers to ensure that each client connects to a nearby server, while inter-server connections are established by a high-speed private network to reduce RTT~\cite{cheng2022are}.

\vspace{0.02in}
\noindent {\bf Protocols.}
When all users wear Vision Pro, FaceTime delivers the content via QUIC~\cite{rfc9000}, different from prior studies~\cite{nistico2020comparative} that reported 
its use of RTP~\cite{rfc3550}.
{However, the transmissions between Vision Pro users and non-Vision Pro users revert to RTP.}
We verify that its Payload Types (PTs) field, which indicates the audio and video codecs~\cite{nistico2020comparative,rfc3551}, remains consistent with that in traditional 2D video calls on FaceTime. 
{This may be because non-Vision Pro users are unable to render spatial personas.}
Thus, Vision Pro pre-renders the spatial persona and delivers it with 2D video.
The other three applications continue to rely on RTP, even when all participants use Vision Pro, probably because their personas are 2D (\S\ref{sec:bakcground}).

\begin{figure*}[t]

  \begin{minipage}[t]{0.32\linewidth}
    \centering  
    \hspace{-0.3in}
    \includegraphics[width=\columnwidth]{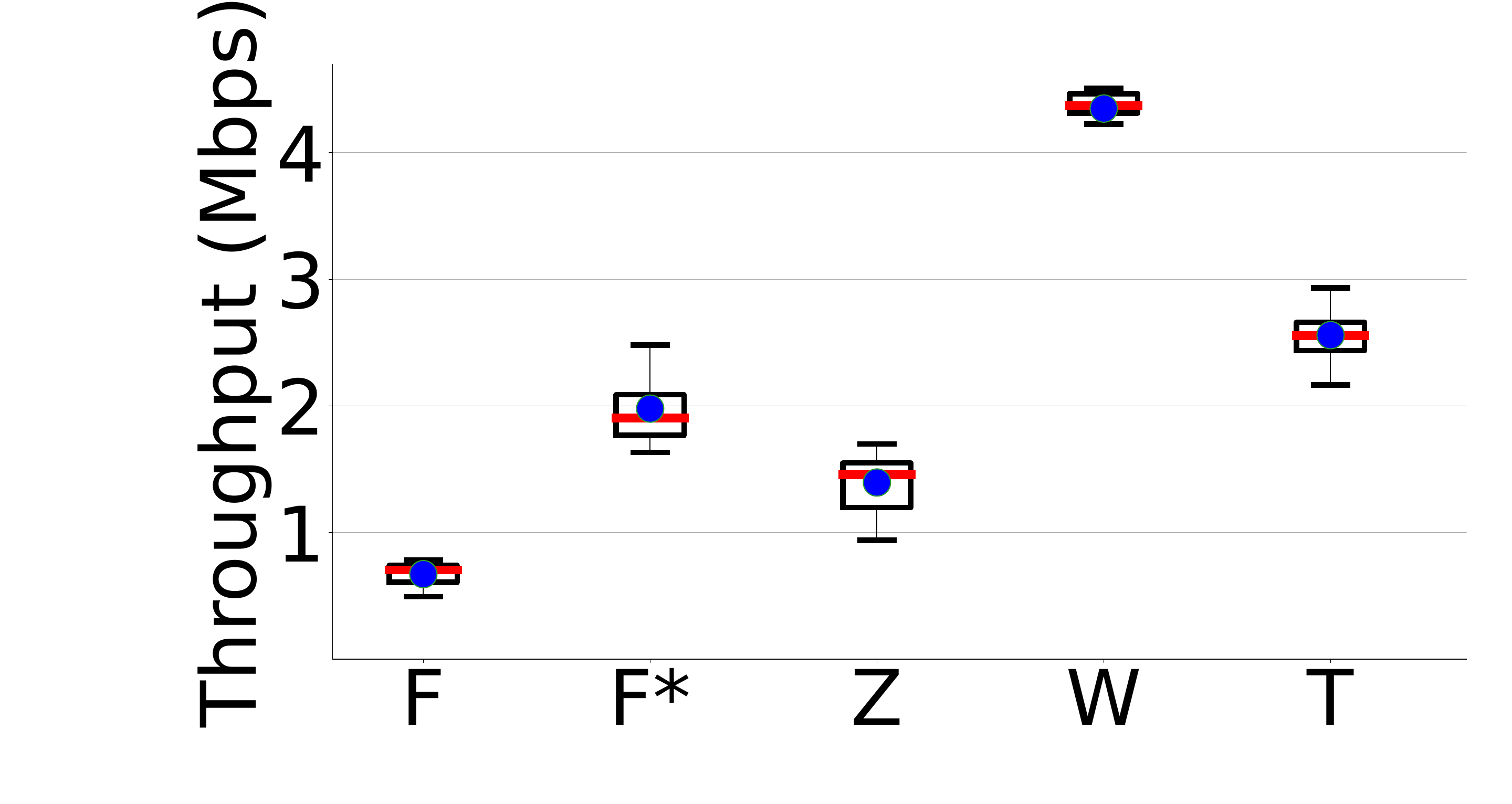} 
    \vspace{-0.15in}
    \caption{ 
    Throughput 
    of FaceTime with spatial persona (F), FaceTime with 
    2D persona (F*), Zoom (Z), Webex (W), and Teams (T) 
    with two participants. Blue dots represent mean values.} 
    \vspace{-0.1in}
    \label{fig:tput_diff_platforms}
    \end{minipage}
    \hspace{0.03in}
    \begin{minipage}[t]{0.66\linewidth}
    \centering    
    \includegraphics[width=0.49\columnwidth]{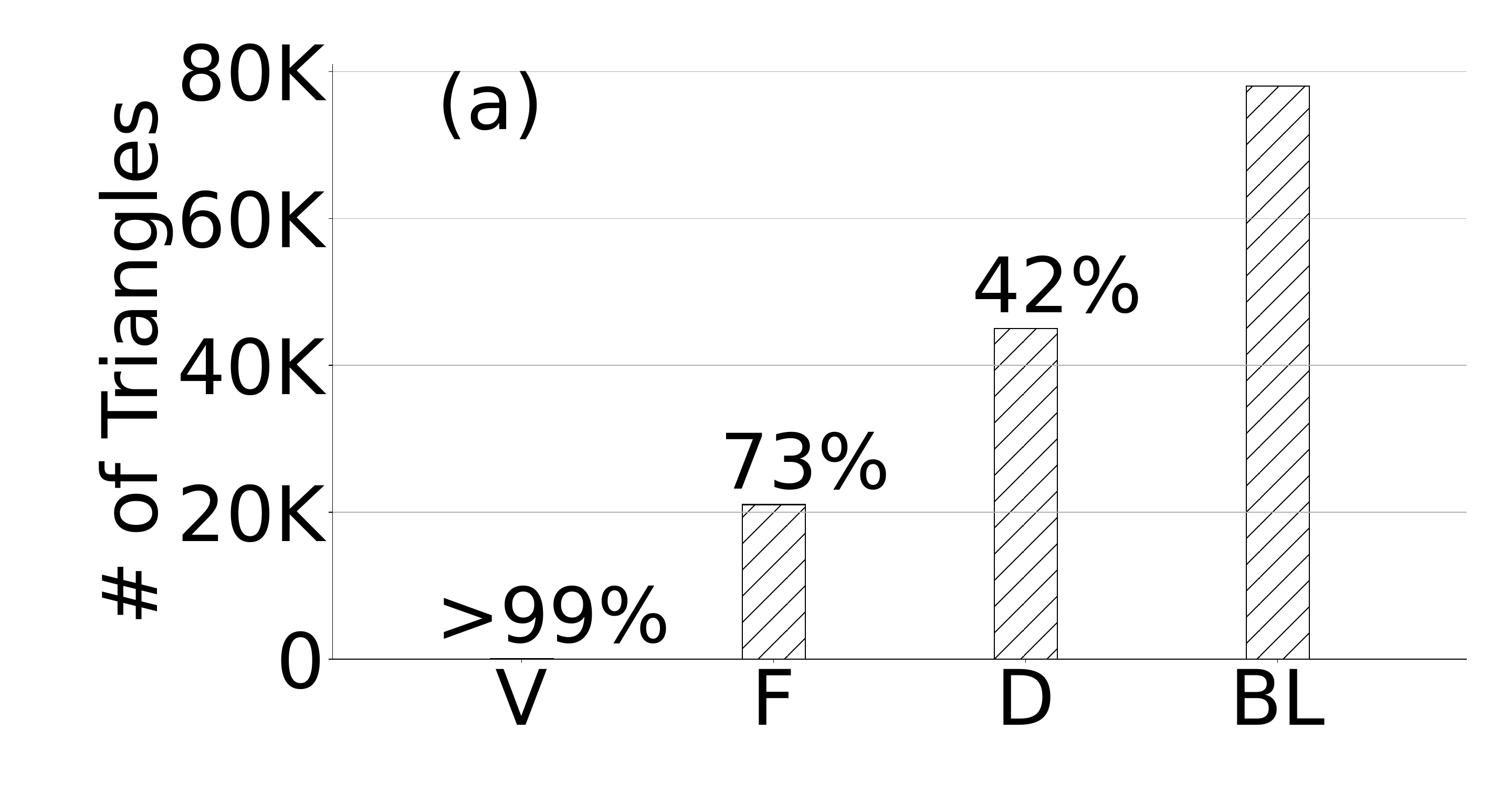} 
    \includegraphics[width=0.49\columnwidth]{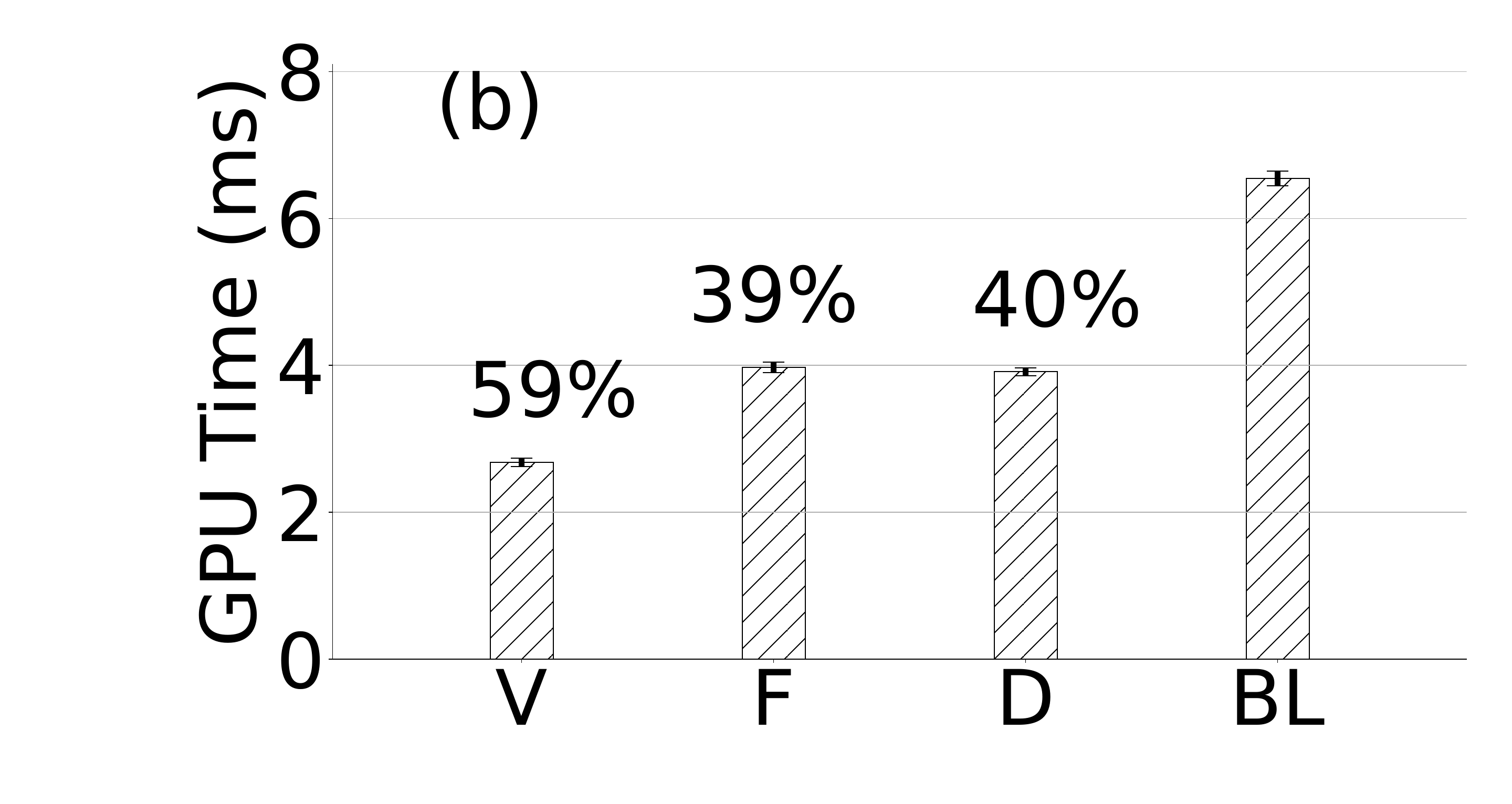} 
    \vspace{-0.15in}
    \caption{Number of triangles (a) and GPU processing time per frame (b) for the rendered spatial persona with various optimizations: viewport adaptation (V), foveated rendering (F), and distance-aware (D). The baseline (BL) is when the user stares at the spatial persona at a close distance of half a meter. The optimizations reduce the number of rendered triangles, leading to decreased GPU rendering time.
} 
    \vspace{-0.1in}
    \label{fig:triangle_gpu}
    \end{minipage}

\end{figure*}

\vspace{-0.05in}
\subsection{Throughput Analysis}
\label{sec:tput}
\vspace{-0.05in}

We next examine the throughput of Vision Pro for the four VCAs.
By analyzing the uplink and downlink traffic of each VCA, we find that their servers are primarily used for data forwarding.
Thus, the throughput of Vision Pro can be considered mainly as the data rate required by the spatial persona.
For FaceTime, we compare the throughput of the spatial persona 
and that of the 2D persona, 
as their underlying protocols~(\S\ref{sec:server_infra}) and types of delivered content~(\S\ref{sec:delivered}) are different.

Figure~\ref{fig:tput_diff_platforms} shows our measurement results for two-user experiments, with the 95th, 75th, 25th, and 5th percentiles, median (red bar), and mean (blue dot).
Surprisingly, the throughput of a spatial persona is the lowest with $\sim$0.7 Mbps, while the throughput of a 2D persona for FaceTime is $\sim$2 Mbps.
Our further analysis~(\S\ref{sec:delivered}) indicates this is because FaceTime employs the semantic communication paradigm~\cite{cheng2023enriching} to optimize bandwidth consumption for spatial personas.
{Among 2D personas for other applications, Webex consumes the highest bandwidth ($>$4 Mbps), while Zoom requires only $\sim$1.5 Mbps.
This is mainly because of their different resolutions for 2D personas (1920$\times$1080 on Webex {\em vs.} 640$\times$360 on Zoom).
Additionally, the different video compression approaches utilized by these applications may affect bandwidth consumption~\cite{nistico2020comparative}.
{Note that while the 2D persona has a background, as shown in Figure~\ref{fig:persona}(b), we observe that it is static and consistent across different applications, suggesting that it does not need to be delivered.
}

\subsection{What is Being Delivered?}
\label{sec:delivered}
Immersive telepresence 
could use three 
approaches for 3D content delivery: 1) direct streaming~\cite{jin2024meshreduce,guan2023metastream}, 2) pre-rendering to 2D video before delivery~\cite{liu2024muv2},
and 3) delivery of semantic information~\cite{cheng2023enriching}.
We next examine which method is used for spatial personas on FaceTime, the only VCA 
currently supporting this feature (\S\ref{sec:bakcground}).

\vspace{0.02in}
\noindent \textit{\textbf{Direct 3D Data Streaming.}} This approach involves sending the 3D model of the spatial persona to the receiver, who then renders it for display. 
Due to the data-hungry nature of 3D data, this approach may consume excessive bandwidth (\eg $>$1 Gbps~\cite{orts2016holoportation}).

{The RealityKit tool~\cite{realitykit} shows that the 3D mesh of a spatial persona consists of 78,030 triangles, representing the complexity of the mesh~\cite{peng2005survey}.
To estimate the bandwidth requirements for streaming the spatial persona on Vision Pro, we select ten different meshes of human heads from Sketchfab~\cite{sketchfab}, with the number of triangles varying from $\sim$70K to $\sim$90K.}
We compress these meshes using Draco~\cite{draco}, a 3D data compression tool widely used in telepresence systems~\cite{guan2023metastream,jin2024meshreduce}, and stream them at 90 FPS, the target frame rate of Vision Pro (\S\ref{sec:testbed}).
We find that the bandwidth consumption is 108.4$\pm$16.7 Mbps, even without texture (\ie the surface details of 3D mesh~\cite{peng2005survey}), drastically higher than $\sim$0.7 Mbps consumed by a spatial persona (Figure~\ref{fig:tput_diff_platforms}).
It follows that the spatial persona is currently not delivered using the 3D mesh format.

\vspace{0.02in}
\noindent \textit{\textbf{Streaming of 
2D Video.}} 
When the delivered content is 2D video, it could be directly captured by the sender 
or 
rendered from the spatial persona of the sender (\eg according to the predicted future viewport of the receiver~\cite{liu2024muv2}).
We find that the content is not the video captured by the sender as a change in the sender's appearance (\eg a sticker on the face) is not communicated to the receiver. 

Next, we investigate if the delivered content is the pre-rendered spatial persona.
We cannot use throughput measurements to determine this, given that we do not have direct access to the resolution of the rendered spatial persona.
A distinguishing characteristic of delivering pre-rendered immersive content is that the display latency between the local real-world objects and the spatial persona at the receiver~(\S\ref{sec:setup}) should be influenced by the network delay.
For example, if the network latency is high and the receiver changes the viewport, it will significantly delay the display of the pre-rendered spatial persona of the sender for the new viewport.

To measure the difference in display latency, we record the content displayed on U1's Vision Pro, which includes both local real-world objects and U2's spatial persona.
We let U1 abruptly change the viewport to observe a different portion of U2's spatial persona from a new angle.
For example, before changing the viewport, U1 views U2's spatial persona from the front, where only its front face is visible.
Then, U1 quickly shifts the viewport to the left, observing one side of U2's spatial persona, including the entire ear.
As U1 changes the viewport, we measure the time taken to render {the newly emerged real-world objects} and U2's spatial persona to determine the difference in display latency.
We use Linux \texttt{tc}~\cite{linux_tc} to introduce extra network delays ranging from 0 to 1,000~ms between U1 and U2. Our experiments indicate that the measured difference in display latency remains consistent (<16 ms), suggesting that the delivered content is not 2D video captured/rendered by the sender.

\vspace{0.02in}
\noindent \textit{\textbf{Delivery of Semantic Information.}} Semantic communication is an emerging content delivery paradigm.
For immersive telepresence, it involves sending only the meaningful semantic data of remote users to the receiver, who then reconstructs the 3D representation (\eg mesh) of remote users using the received data~\cite{cheng2023enriching}. 

For the human body, keypoints represent a primary choice for conveying semantic information~\cite{cheng2023enriching}.
Given that the spatial persona primarily includes the head and hands (Figure~\ref{fig:persona}), we explore the bandwidth requirement for delivering keypoints in these areas to verify whether semantic communication is the used method.
Specifically, we utilize a ZED 2i RGB-D camera~\cite{zed2i} to capture a video 
of 2,000 frames containing the head and hand regions of the user.
We employ the widely used 68 facial keypoints from \textit{dlib}~\cite{king2009dlib} and 21 hand keypoints from \textit{OpenPose}~\cite{cao2017realtime}.
As the spatial persona primarily tracks the eye and mouth areas for facial expressions, as well as hand movements, we compress the 32 (mouth \& eyes) $+$ $2 \times 21$ (hands) $= 74$ extracted keypoints using LZMA~\cite{gupta2017modern} and stream them at 90 FPS.
The average throughput is 0.64$\pm$0.02 Mbps, close to the bandwidth consumed by a spatial persona (0.67 Mbps on average).
This suggests that FaceTime utilizes semantic communication to optimize bandwidth consumption for spatial personas.

Although semantic communication consumes less bandwidth than 2D/3D content streaming, it leads to challenges in supporting rate adaptation. 
We conduct experiments by using Linux \texttt{tc}~\cite{linux_tc} to constrain the bandwidth.
When the uplink bandwidth is 0.7 Mbps, the spatial persona becomes unavailable, with ``poor connection'' displayed on the screen.
This may be because semantic data must be fully delivered for successful content reconstruction~\cite{cheng2023enriching}.
While currently, the bandwidth consumption for spatial personas on FaceTime is relatively low, rate adaption may still be necessary.
The spatial persona does not yet provide a fully immersive experience, such as a photorealistic representation.
At present, it captures only the head and hands and relies on an avatar-based model rather than lifelike representations of actual users (Figure~\ref{fig:persona}). 
Achieving a high-quality, full-body immersive representation could demand significantly higher throughput than what we have observed~\cite{cheng2022are}.

\vspace{0.02in}
\noindent{\bf Implications~\circled{2}~:}
We identify that the spatial persona on FaceTime utilizes semantic communication to reduce bandwidth consumption.
{However, semantic communication is not a silver bullet for immersive telepresence and
has its own technical challenges.}
For example, since it requires all semantic data to be successfully delivered for reconstruction~\cite{cheng2023enriching}, semantic communication is not resilient to data loss and makes rate adaptation challenging. 
Additionally, as semantic information is inherently sparse, the reconstructed content may suffer from a loss of fidelity.
Conversely, other streaming approaches, such as direct 3D data streaming and 2D video streaming, have their own limitations, as discussed earlier.
{A potential solution is to have servers intelligently select different streaming approaches for each client based on their available network and computational resources~\cite{liu2024muv2}.} %

\begin{figure*}[t]
    \centering   
    \hspace{-0.2in}
    \includegraphics[width=0.33\linewidth]{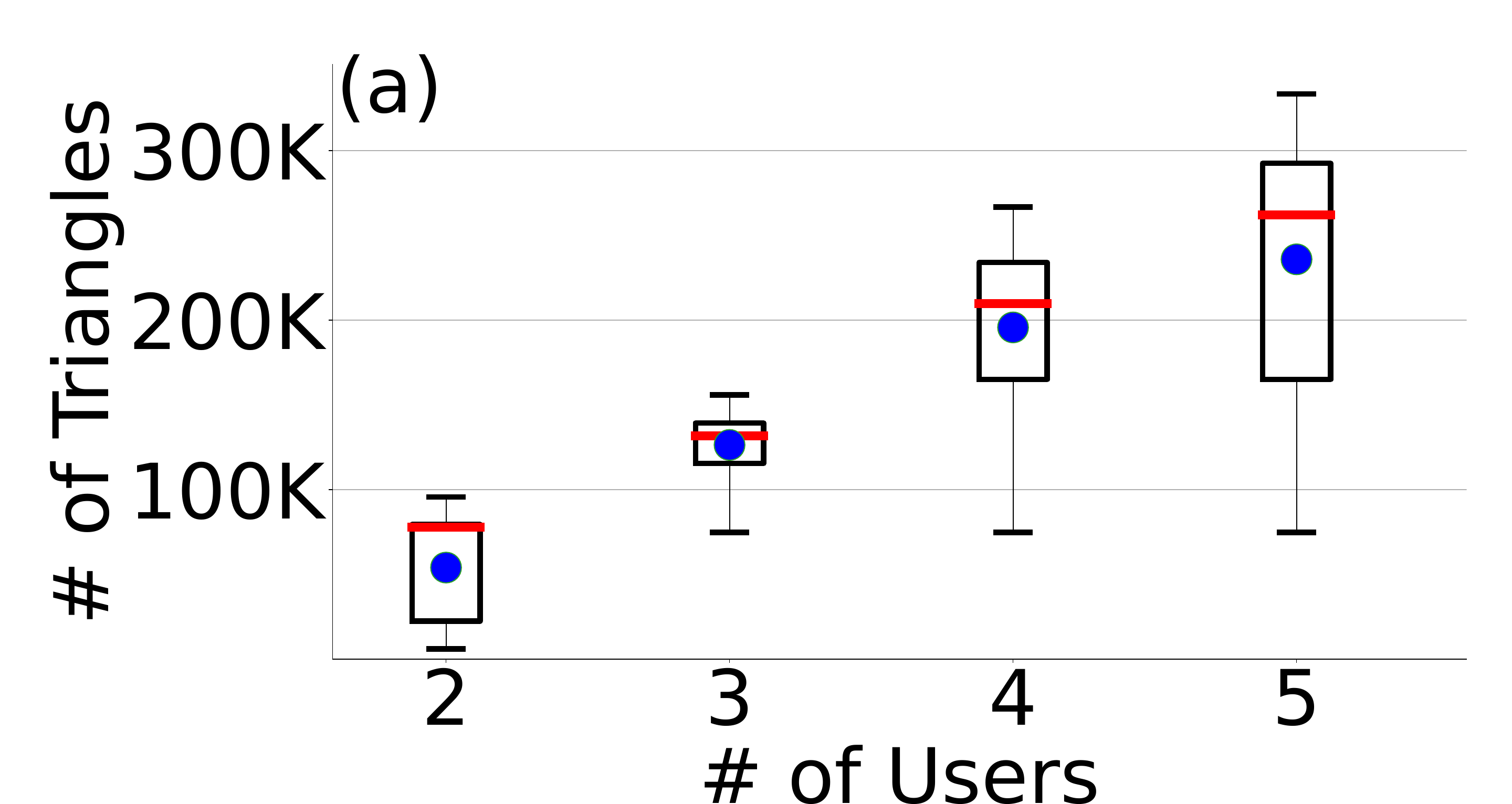}   
    \includegraphics[width=0.33\linewidth]{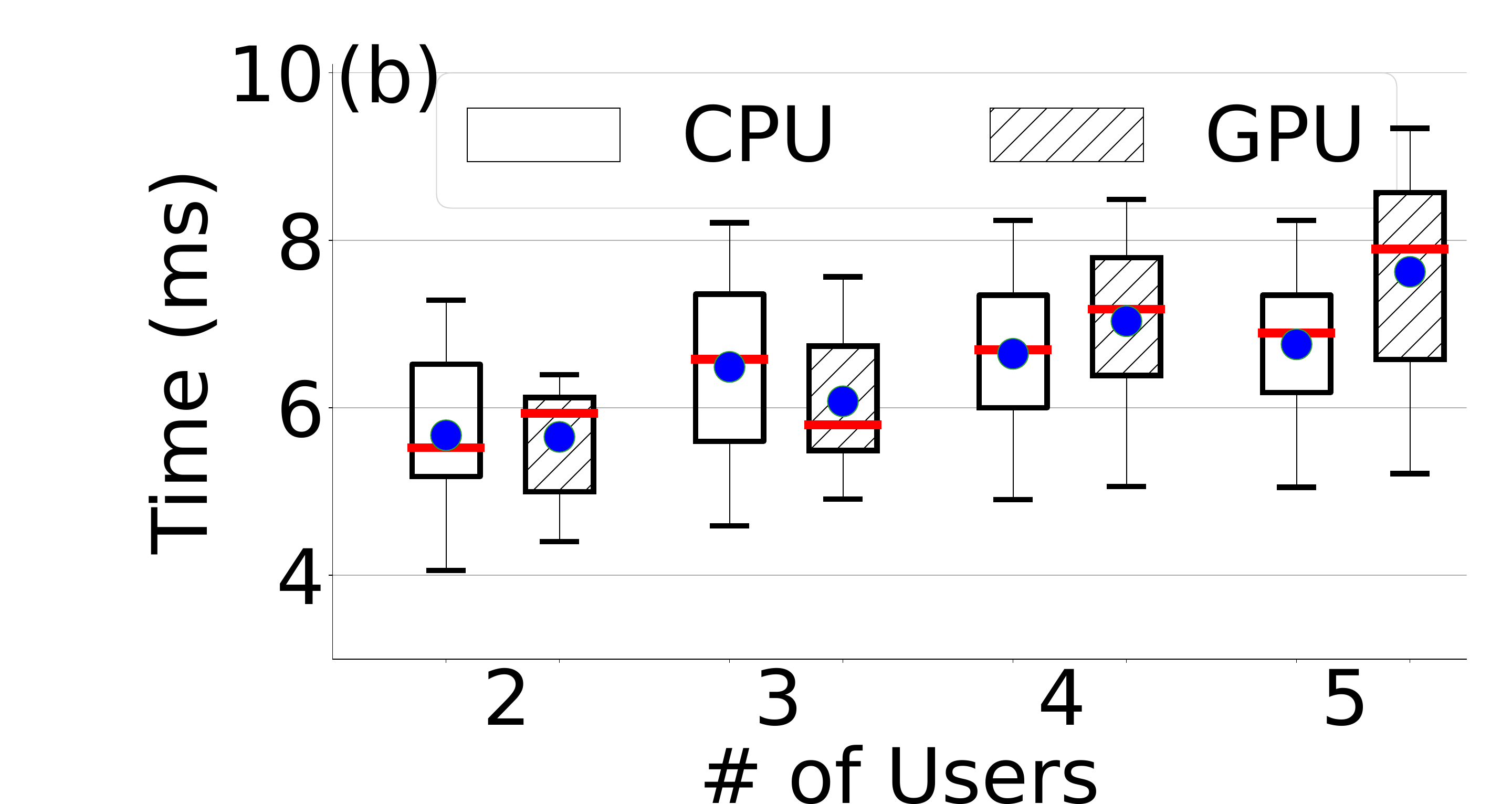}  
    \includegraphics[width=0.33\linewidth]{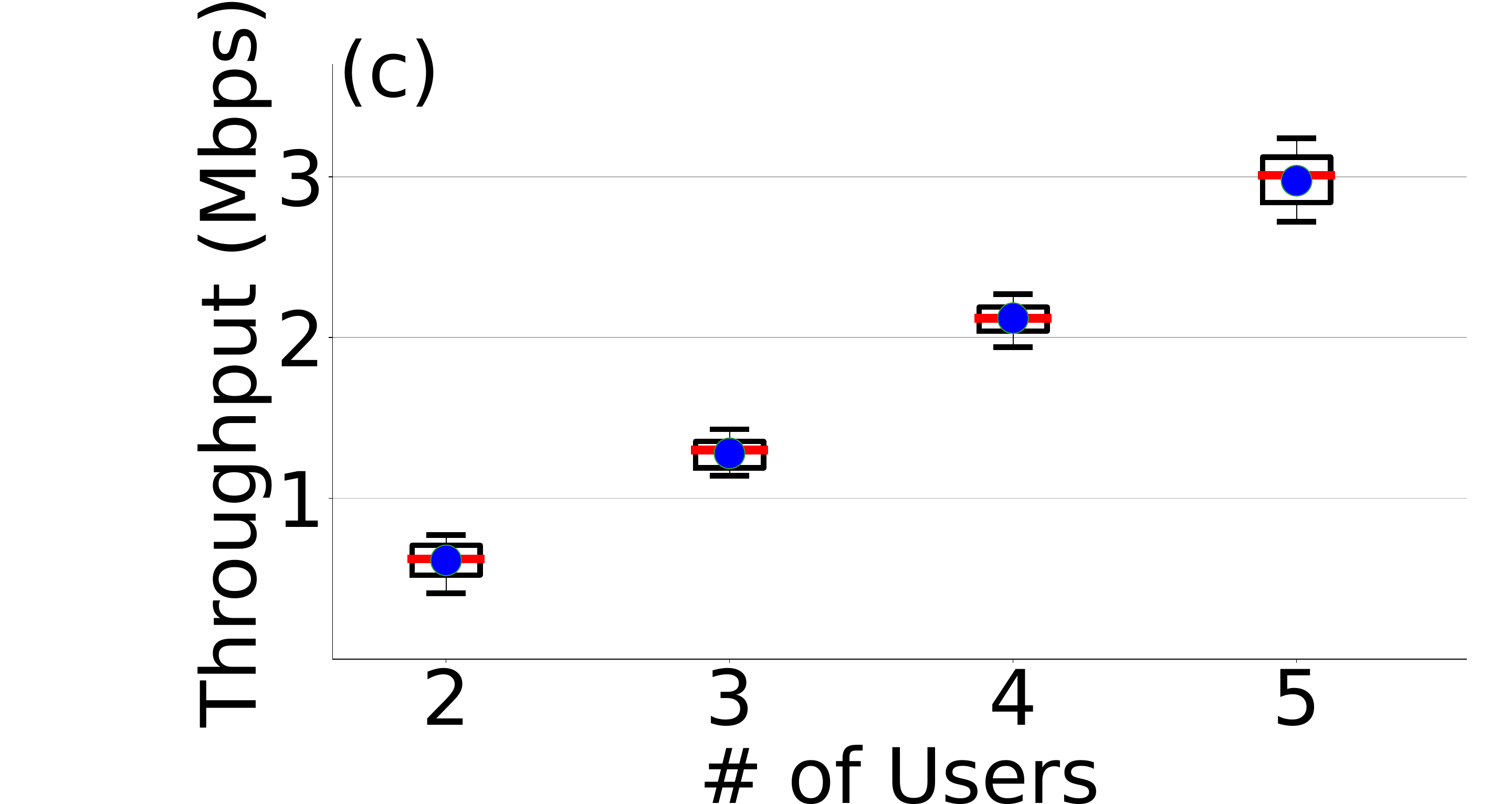}   
    \vspace{-0.05in}
    \caption{Number of rendered triangles (a), CPU/GPU processing time (b), and downlink throughput (c) of spatial personas for FaceTime with the number of users varying from 2 to 5.} 
    \label{fig:scalability}
    \vspace{-0.10in}
\end{figure*}

\subsection{Visibility-aware Optimization}
\label{sec:visibility}

Visibility-aware optimizations can drastically reduce communication and computing overhead in immersive video streaming~\cite{qian2018flare,han2020vivo,wu2024theia}. 
However, there is limited research on their adoption in commercial products. 
To fill this critical gap, we investigate the potential deployment of various visibility-aware optimizations for spatial personas on FaceTime.
We analyze the number of triangles of rendered meshes for a spatial persona, indicative of its visual quality~(\S\ref{sec:testbed}), along with CPU/GPU processing time and bandwidth consumption. 
As a baseline, we consider U1 viewing U2’s spatial persona from the closest distance (approximately half a meter), at which U2's entire persona just fits within U1's screen. In this scenario, no visibility-aware optimization should be applied.

We experiment with the spatial personas for the following possible optimizations: 1) viewport adaptation, 2) foveated rendering, 3) distance-aware optimization, and 4) occlusion-aware optimization. 
We find that the first three optimizations are employed to reduce the number of 
rendered triangles and thus decrease GPU processing time, as shown in Figure~\ref{fig:triangle_gpu}.
Next, we will detail our conducted experiments and discuss the potential for further optimizations.

\vspace{0.02in}
\noindent {\bf Viewport Adaptation} processes only content in the user's viewport~\cite{qian2018flare,han2020vivo}.
We verify whether Vision Pro adopts it for spatial persona by having U1 turn the head to make U2's spatial persona out of U1's viewport.
Our results show a decrease in the number of rendered triangles, from 78,030 to 36, and a 59\% reduction in GPU rendering time per frame, from 6.55$\pm$0.11 ms to 2.68$\pm$0.05 ms.

\vspace{0.02in}
\noindent {\bf Foveated Rendering}
benefits from the human visual system~\cite{van1996perceptual} to render with the highest visual quality for only foveal content around the center of the eye gaze and lowers the quality toward the periphery~\cite{guenter2012foveated}. 
In our setup, U2’s spatial persona appears at the left corner of U1’s viewport when U1 gazes toward the right corner, placing U2's persona in U1's peripheral vision. This results in a 73\% reduction in the number of rendered triangles (21,036), and a 39\% decrease in GPU rendering time per frame (3.97$\pm$0.07 ms).

\vspace{0.02in}
\noindent {\bf Distance-aware Optimization} adjusts the rendered 3D content based on viewing distance~\cite{han2020vivo}. 
We vary the viewing distance from half a meter to ten meters in increments of half a meter.
Beyond three meters, a lower quality spatial persona is displayed, with the number of rendered triangles reduced by 42\% to 45,036, and the GPU rendering time per frame reduced by 40\% to 3.91$\pm$0.05 ms.

\vspace{0.02in}
\noindent {\bf Occlusion-aware Optimization} reduces the quality or omits the rendering of occluded content~\cite{han2020vivo}.
We experiment with five Vision Pro users, U1 through U5, and arrange U2 to U5 in a line, with U1 observing the rest from the front. If occlusion-aware optimization is implemented, the spatial personas of U3 to U5 should not be rendered on U1's Vision Pro, as they are occluded by U2. However, compared to the case where all users are visible, we do not observe a reduction in the number of rendered triangles and GPU processing time for U1, indicating that this optimization is not adopted.

Despite the adoption of several visibility-aware optimizations by Vision Pro for spatial persona, this does not translate into a reduction in bandwidth consumption and CPU processing time compared to scenarios without these optimizations.
It suggests that optimizations are applied solely at the rendering stage but not during content delivery. 
The lack of bandwidth optimization might explain why the CPU processing time remains unchanged, since the CPU on Vision Pro is tasked with processing the received data, as indicated by the RealityKit tool~\cite{realitykit}.

\vspace{0.02in}
\noindent{\bf Implications~\circled{3}~:}
Our measurements indicate that FaceTime employs several visibility-aware optimizations to reduce computational overhead for spatial personas. 
However, it has not yet implemented occlusion-aware optimizations, which could be beneficial when multiple users and/or objects are present within the same scene.
Moreover, {these visibility-aware optimizations do not benefit the data transmission stage}. 
Nevertheless, implementing such optimizations to reduce bandwidth consumption is feasible. For example, if the content is known to fall outside of a receiver’s viewport, it could be omitted from delivery to conserve bandwidth~\cite{qian2018flare,han2020vivo,wu2024theia}. 
These optimizations could be further applied in immersive telepresence systems to {reduce bandwidth consumption}.

\subsection{Scalability Analysis}
\label{sec:scalability}
We finally investigate the scalability of spatial personas on FaceTime by measuring the throughput and rendering overhead as the number of users increases.
Specifically, we have at most five Vision Pro users joining 
a telepresence session, the maximum number currently supported by FaceTime~\cite{spatial_persona}.
The available bandwidth for each user is at least 100~Mbps.
Figure~\ref{fig:scalability} shows the number of rendered triangles, CPU/GPU processing time, and downlink throughput as a function of the number of concurrent users.

Although increasing the number of spatial personas almost linearly raises the average number of rendered triangles, the 5th percentile for five users remains almost the same as that for three users, as shown in Figure~\ref{fig:scalability}(a).
This can be attributed to the visibility-aware optimizations adopted by FaceTime (\S\ref{sec:visibility}).
For instance, as the number of spatial personas increases, some of them may appear in the peripheral regions of the visual field, which will be displayed as a low-quality mesh with few triangles due to foveated rendering.}

Despite the implementation of various visibility-aware optimizations, the GPU rendering time still increases by an average of 34.9\% from two users (5.65$\pm$0.69 ms) to five users (7.62$\pm$1.29 ms), {with the 95th percentile $>$9 ms,} as shown in Figure~\ref{fig:scalability}(b), which is close to the rendering deadline (\ie $\sim$11 ms for 90 FPS).
This likely explains why FaceTime currently supports a maximum of five spatial personas. 
{We also observe the CPU processing time increases by an average of 19.2\% from two users (5.67$\pm$0.69 ms) to five users (6.76$\pm$1.29 ms).}
Figure~\ref{fig:scalability}(c) reveals that the downlink throughput of spatial personas almost linearly increases with the number of users.
This is because the server just simply forwards the data (\S\ref{sec:server_infra}).

\vspace{0.02in}
\noindent{\bf Implications~\circled{4}~:}
The scalability issues related to {resource utilization and bandwidth consumption} for spatial personas on FaceTime significantly impede its ability to support a large number of participants.
A potential solution to address such scalability issues is to utilize remote rendering by offloading the GPU-intensive rendering process to the cloud~\cite{cheng2022are}.
By having the server handle rendering, even with many concurrent users, the server can render them into a 2D video frame.
This ensures that the transmitted data remains independent of user numbers, mitigating scalability issues.

\section{Discussion}
\label{sec:discussion}
\noindent {\bf Fully-automated Measurement Experiments.}
To the best of our knowledge, no existing tool can automatically play back predefined user inputs on Vision Pro.
Thus, we resort to manual experiments in this study.
A potential method for automating experiments is to attach Vision Pro to a robotic arm~\cite{zhou2017measurement}.
However, this may cause the spatial persona not to function, as it needs to track users' facial changes.
We plan to build open-source tools for Vision Pro to facilitate automated and large-scale 
crowd-sourced measurement experiments in the wild.

\vspace{0.02in}
\noindent {\bf Content Decryption.}
To know exactly the delivered content for a spatial persona, a promising solution is to 
decrypt the 
content.
However, FaceTime utilizes QUIC~\cite{rfc9000} to deliver spatial persona~(\S\ref{sec:server_infra}), which is encrypted by TLS 1.3~\cite{zhang2024quic}.
As spatial persona is end-to-end encrypted~\cite{visionpro_privacy}, simply utilizing the man-in-the-middle attack cannot get the TLS certificate, and thus it is challenging to decrypt the content.
Instead of relying on content decryption, analyzing IP headers~\cite{sharma2023estimating} and packet transmission patterns~\cite{michel2022enabling} may help better understand the delivered content for spatial persona.

\vspace{0.02in}
\noindent {\bf Other Use Cases.}
This paper focuses on immersive telepresence, a major use case of remote collaboration.  
Vision Pro also facilitates other use cases such as collaborative whiteboards~\cite{visiopro_freedom} and shared entertainment experiences (\eg playing games and watching movies)~\cite{visiopro_shareplay}, which we plan to explore in the future.

\section{Related Work}
\label{sec:related}
\noindent {\bf Network Measurements on VCAs.}
In recent years, there has been a growing 
research interest in measuring the network performance of VCAs~\cite{nistico2020comparative, chang2021can, MacMillan2021VCAs, michel2022enabling, varvello2022performance, sharma2023estimating, jansen2018performance, he2023measurement}. 
For example, Varvello~\etal~\cite{varvello2022performance} build a large-scale testbed to facilitate the evaluation of videoconferencing performance in the wild.
Sharma~\etal~\cite{sharma2023estimating} utilize IP/UDP headers for QoE estimation of VCAs.
In this paper, we measure the performance of immersive telepresence with these VCAs on Apple Vision Pro.

\vspace{0.02in}
\noindent {\bf Measurement of Immersive Applications.}
Existing studies on the performance of immersive applications have focused on immersive video streaming~\cite{zhou2017measurement}, Web-based extended reality (XR)~\cite{liu2023demystifying}, and social virtual reality (VR) platforms~\cite{cheng2022are, cheng2022reality, lyu2024metavradar, cheng2022will}. 
For instance, MetaVRadar~\cite{lyu2024metavradar} correlates the network traffic of social VR applications with user activities. 
Liu~\etal~\cite{liu2023demystifying} investigate Web-based XR platforms accelerated by WebAssembly~\cite{haas2017bringing}.
This paper measures spatial personas that improve telepresence experiences.

\vspace{0.02in}
\noindent {\bf Telepresence Systems}
are increasingly gaining
attention in the industry (\eg Holoportation~\cite{orts2016holoportation} from Microsoft, Project Starline~\cite{lawrence2021project} from Google, and Codec Avatar~\cite{ma2021pixel} from Meta) and academia (\eg MetaStream~\cite{guan2023metastream}, FarfetchFusion~\cite{lee2023farfetchfusion}, and MeshReduce~\cite{jin2024meshreduce}).
Moreover, the human-computer interaction community has developed in-lab prototypes for specific use cases, such as conducting remote surgeries~\cite{gasques2021artemis}
and teaching physical tasks~\cite{thoravi2019loki}.
In this paper, we measure commercial telepresence systems on Apple Vision Pro.

\section{Conclusion}
\label{sec:conclusion}
This paper presents a first-of-its-kind in-depth and empirical measurement study of immersive telepresence 
on Apple Vision Pro.
Driven by the counter-intuitive results that the required bandwidth of the immersive spatial persona is even lower than its 2D counterpart, we conduct a comprehensive analysis of the 
delivered content. 
{We find that spatial personas utilize semantic communication to optimize bandwidth consumption, which, however, leads to challenges for employing rate adaptation.}
Moreover, we dissect the 
visibility-aware optimizations and the scalability issue of spatial persona.
We hope that our findings can shed light on the design practices of emerging immersive telepresence systems.

\section*{Acknowledgment}
We thank the anonymous reviewers and our shepherd for their insightful feedback, as well as the participants of our experiments for their valuable contributions. This work was partially supported by the National Science Foundation under Grants CNS-2007153, CNS-2212296, and CNS-2235049.

\clearpage


\end{document}